\documentclass[twocolumn]{aastex631}
\newcommand\w{0.49}
\newcommand{\add}[1]{#1}
\newcommand{\addd}[1]{#1}
\newcommand{\red}[1]{#1}
\newcommand{\blue}[1]{#1}

\begin{document}

Accepted for publication in {\it Astrophysical Journal Letters}\newline
\title{Ground-based Observations of Temporal Variation of Cosmic Ray Spectrum during Forbush Decreases}



\author[0000-0002-3776-072X]{W. Mitthumsiri}
\affiliation{Department of Physics, Faculty of Science, Mahidol University, Bangkok 10400, Thailand}

\author[0000-0003-3414-9666]{D. Ruffolo}
\affiliation{Department of Physics, Faculty of Science, Mahidol University, Bangkok 10400, Thailand}

\author[0000-0002-2131-4100]{K. Munakata}
\affiliation{Department of Physics, Shinshu University, Matsumoto, Nagano 390-8621, Japan}

\author[0000-0002-3948-3666]{M. Kozai}
\affiliation{Polar Environment Data Science Center, Joint Support-Center for Data Science Research, Research Organization of Information and Systems, Tachikawa, Japan}

\author[0000-0002-0890-0607]{Y. Hayashi}
\affiliation{Department of Physics, Shinshu University, Matsumoto, Nagano 390-8621, Japan}

\author[0000-0002-4913-8225]{C. Kato}
\affiliation{Department of Physics, Shinshu University, Matsumoto, Nagano 390-8621, Japan}

\author{P. Muangha}
\affiliation{Department of Physics, Faculty of Science, Mahidol University, Bangkok 10400, Thailand}

\author[0000-0001-7771-4341]{A. S\'aiz}
\affiliation{Department of Physics, Faculty of Science, Mahidol University, Bangkok 10400, Thailand}

\author[0000-0001-7929-810X]{P. Evenson}
\affiliation{Bartol Research Institute, Department of Physics and Astronomy, University of Delaware, Newark, DE, USA}

\author[0000-0003-4865-6968]{P.-S. Mangeard}
\affiliation{Bartol Research Institute, Department of Physics and Astronomy, University of Delaware, Newark, DE, USA}

\author[0000-0001-5681-6883]{J. Clem}
\affiliation{Bartol Research Institute, Department of Physics and Astronomy, University of Delaware, Newark, DE, USA}

\author[0000-0003-3272-6896]{S. Seunarine}
\affiliation{Department of Physics, University of Wisconsin
River Falls, River Falls, WI 54022, USA}

\author[0000-0002-1664-5845]{W. Nuntiyakul}
\affiliation{Department of Physics and Materials Science, Faculty of Science, Chiang Mai University, Chiang Mai 50200, Thailand}

\author[0009-0006-3569-7380]{N. Miyashita}
\affiliation{Department of Physics, Shinshu University, Matsumoto, Nagano 390-8621, Japan}

\author[0000-0001-9400-1765]{R. Kataoka}
\affiliation{Science and Technology Group, 
Okinawa Institute of Science and Technology, Okinawa, Japan}

\author[0000-0002-6105-9562]{A. Kadokura}
\affiliation{Polar Environment Data Science Center, Joint Support-Center for Data Science Research, Research Organization of Information and Systems, Tachikawa, Japan}

\affiliation{National Institute of Polar Research, Tachikawa, Japan}

 \author[0000-0002-3067-655X]{S. Miyake}
 \affiliation{National Institute of Technology (KOSEN), Gifu College, Gifu, Japan}

\author[0000-0002-2464-5212]{K. Iwai}
\affiliation{Institute for Space-Earth Environmental Research, Nagoya University, Nagoya, Japan}

\author[0000-0001-8466-1938]{H. Menjo}
\affiliation{Institute for Space-Earth Environmental Research, Nagoya University, Nagoya, Japan}

\author[0000-0002-8351-6779]{E. Echer}
\affiliation{National Institute for Space Research, São José dos Campos, Brazil}

\author[0000-0002-4361-6492]{A. Dal Lago}
\affiliation{National Institute for Space Research, São José dos Campos, Brazil}

\author[0000-0002-9737-9429]{M. Rockenbach}
\affiliation{National Institute for Space Research, São José dos Campos, Brazil}

\author[0000-0002-7720-6491]{N. J. Schuch}
\affiliation{Southern Space Coordination, National Institute for Space Research, Santa Maria-RS, Brazil}

\author[0000-0003-2931-8488]{J. V. Bageston}
\affiliation{Southern Space Coordination, National Institute for Space Research, Santa Maria-RS, Brazil}

\author[0000-0003-1485-9564]{C. R. Braga}
\affiliation{The Johns Hopkins University Applied Physics Laboratory, Laurel, MD, USA}

\author[0000-0002-4486-237X]{H. K. Al Jassar}
\affiliation{Physics Department, Kuwait University, Kuwait City, Kuwait}

\author[0000-0002-2090-351X]{M. M. Sharma}
\affiliation{Physics Department, Kuwait University, Kuwait City, Kuwait}

\author[0000-0003-3691-6806]{N. Burahmah}
\affiliation{Department of Physics, College of Science, Kuwait University, Sabah AlSalem University City, Kuwait}

\author[0000-0002-4807-4258]{F. Zaman}
\affiliation{Department of Physics, College of Science, Kuwait University, Sabah AlSalem University City, Kuwait}

\author[0000-0001-7463-8267]{M. L. Duldig}
\affiliation{School of Natural Sciences, University of Tasmania, Hobart, Australia}


\author[0000-0002-7037-322X]{I. Sabbah}
\affiliation{Department of Applied Sciences, College of Technological Studies, Public Authority for Applied Education and Training, Shuwaikh, Kuwait}

\author[0000-0001-6584-9054]{T. Kuwabara}
\affiliation{Bartol Research Institute, Department of Physics and Astronomy, University of Delaware, Newark, DE, USA}

\correspondingauthor{D. Ruffolo, K. Munakata}
\email{ruffolo.physics@gmail.com, kmuna00@shinshu-u.ac.jp}



\begin{abstract}

Observations of temporary Forbush decreases (FDs) in the Galactic cosmic ray (GCR) flux due to passage of solar storms are useful for space weather studies and alerts. \blue{Here we introduce techniques that use} global networks of ground-based neutron monitors and muon detectors \blue{to measure} variations of GCR rigidity spectra in space \blue{during FDs} by: A) fitting count rate decreases 
\addd{for power-law rigidity spectra in space}
with anisotropy up to second order, and B) using the “leader fraction” derived from a single neutron monitor. 
\red{We}
\blue{demonstrate}
\red{that both provide consistent results for hourly spectral index variations for five major FDs and they agree with daily space-based data when available from AMS-02. We have also made the neutron monitor leader fraction publicly available in real time. This work verifies that ground-based observations can be used to precisely monitor GCR spectral variation over a wide range of rigidities during space weather events, with results in real time or from short-term post-analysis.}

\end{abstract}



\section{Introduction} \label{sec:intro}

Solar flares and interplanetary coronal mass ejections (ICMEs)
have various impacts on human activity known as space weather effects \citep{Knipp11}.  
Among the numerous techniques used to provide warnings of ongoing space weather events, one avenue focuses on effects of interplanetary shocks and their ICME drivers on Galactic cosmic rays (GCRs),  
including temporary Forbush decreases (FDs) \citep{Forbush37,Cane00} (for an example of an automated FD warning system, see \url{https://www.sws.bom.gov.au/Geophysical/1/4}).
A related 
method watches
for an anisotropic precursory decrease before the arrival of an interplanetary shock, i.e., a loss-cone decrease in GCR ions from certain directions before a Forbush decrease, which can in principle provide advance warning up to 12 hours before certain space weather effects \citep{NagashimaEA92,BelovEA95,Ruffolo99,Marlos14,LeerungnavaratEA03,PapaioannouEA09}.

Real-time space weather warning systems based on cosmic ray ions commonly use ground-based detectors of atmospheric secondary particles, monitoring the count rate due to the detector response over a wide range of primary cosmic ray energies.
The most commonly used detectors are neutron monitors (NMs) \citep{Simpson48,NM64}, which can provide stable monitoring of count rates over time scales up to decades \citep{Usoskin05}.
However, precise determination of short-term variation in the GCR energy (or rigidity) spectrum has been more challenging; for example, a comparison between count rates of NMs with different geomagnetic cutoffs (thresholds) may not correctly indicate short-term changes in the spectral index, presumably due to local atmospheric effects on the count rates \citep{Ruffolo16}.
Therefore, techniques have been developed using specially designed electronics to record the distribution of time delays from one neutron arrival to the next in an NM \citep{Bieber04-mult}, and from this distribution to extract the leader fraction \citep{Ruffolo16} (inverse neutron multiplicity) as a measure of the cosmic ray spectral index $\mathit{\Gamma}$ (i.e., negative logarithmic slope of the cosmic intensity vs.\ rigidity $P=pc/q$, where $p$ is the particle's momentum and $q$ is its charge).
Measuring the spectral index at a single NM avoids systematic errors from local atmospheric effects on the count rate, which are effectively divided out when calculating the leader fraction.

GCR intensity variation observed at ground level generally consists of two components: 1) temporal variation in GCR density (or the isotropic component of intensity) and 2) variation because of GCR directional anisotropy, as the detector's look direction changes with Earth's rotation. To analyze these components separately and accurately requires global detector networks that simultaneously monitor the whole sky around Earth.

NMs, which detect atmospheric secondary neutrons, have a maximum response to primary GCRs with median rigidities between 10 GV and 35 GV. 
An NM is an omnidirectional detector, with a response dominated by nearly vertically incident GCRs, so anisotropy information is obtained either from a single detector, as Earth rotates \citep{Elliot51}, or over faster timescales by comparing data from NMs with different look directions \citep{Altukhov70,Bieber98}. 

Muon detectors (MDs) detect muons produced through the hadronic interaction between GCRs and atmospheric nuclei. Because a higher primary energy is needed to produce muons with sufficient Lorentz factor and relativistic time dilation to reach ground level before decaying, muon detectors (MDs) have a response to primary GCRs with higher median rigidities, between $\sim$50 and $\sim$100 GV. 
A single MD can have numerous directional channels because the incident directions of muons closely represent the incident directions of primary GCRs at the top of the atmosphere. 
The Global Muon Detector Network (GMDN) was established in 2006 with four multidirectional surface muon detectors at Nagoya (Japan), Hobart (Australia), Kuwait City (Kuwait), and São Martinho (Brazil) \citep{Marlos14}. \add{The global fit analysis of data from networks of NMs and MDs worldwide allows us to accurately separate anisotropy from spectral effects \citep{muna22}, and deduce the GCR rigidity spectrum over a wide rigidity range. 
Such information can only be derived from the cosmic ray measurements and
observationally constrains the three-dimensional macroscopic picture of the interaction between coronal
mass ejections and the ambient solar wind, which is essential for prediction of large magnetic storms \citep{Kihara21}.}

Daily fluxes of protons (which dominate GCR ions) from the Alpha Magnetic Spectrometer (AMS-02) experiment on board the International Space Station \add{(ISS)} are available for 2011-2019, during the previous sunspot cycle \citep{AMS21-dailyprotons}. 
\citet{wang23} recently analyzed these data to determine GCR spectral variation during a large number of FDs.
\add{The AMS-02 detector can identify the species and rigidity of charged cosmic rays with a large geometric factor \citep{AMS21-7years}.
It orbits Earth along with ISS, with a period of 93 min and orbital inclination of 52$^\circ$.
}
While the viewing direction 
\add{and geomagnetic cutoff (threshold) rigidity}
of the detector rapidly \add{change} during its orbit and \add{it} does not evenly sample the entire sky, the daily averages provide a reasonable estimate of the omnidirectional cosmic ray spectrum.  

Here we present the temporal variation of the GCR rigidity spectral index $\mathit{\Gamma}$ measured from the South Pole leader fraction and from global fit analysis of NM and MD count rates during five FDs, including comparison with the AMS-02 data that are available for three of the FDs.

\section{Data and Analysis Methods} \label{sec:methods}

\subsection{AMS-02 data}

We use the daily proton rigidity spectrum measured by AMS-02 from 20 May 2011 to 29 October 2019 \citep{AMS21-dailyprotons}. The proton flux $\Phi_k$ given in each of 30 rigidity bins is assigned to the geometric mean $P_k$ of minimum and maximum rigidities of the $k$-th rigidity bin, with $P_k$ values ranging from 1.10 GV to 83.50 GV. We calculate the fractional change in GCR density at $P_k$ as $\Delta I_k=(\Phi_k-\Phi^0_k)/\Phi^0_k$, where $\Phi^0_k$ is the average of $\Phi_k$ over a reference interval of the initial three days, before the FD onset. The error of $\Delta I_k$ is calculated from the total systematic error of $\Phi_k$ given in the AMS-02 data table. 
We then estimate the local power-law index $\mathit{\Gamma}_k$ as the negative slope of a weighted least-squares  linear fit to $\log\Phi_i$ versus $\log P_i$ for $i=k-3,...,k+3$, i.e., for the 7 data values closest to the $P_k$ of interest.  
In this work we estimate $\mathit{\Gamma}_{10.5}$ for the rigidity $P_k=10.5$ GV closest to the South Pole NM median primary rigidity of 11.3 GV. \add{$\Delta\mathit{\Gamma}_{10.5}$ is then calculated as $\mathit{\Gamma}_{10.5}-\mathit{\Gamma}^0_{\rm 10.5}$ where $\mathit{\Gamma}^0_{\rm 10.5}$ is the average of $\mathit{\Gamma}_{10.5}$ over the reference interval.}

\subsection{Global Fit Analysis (GFA)}

GFA deduces the best-fit model GCR intensity in space using 12 independent parameters representing GCR density and anisotropy for each hour that reproduce hourly count rates recorded by NMs and MDs with the minimum chi-square \citep{muna22,muna24}. In this work, we \add{simultaneously fit} 85 hourly count rates recorded by 25 NMs and 60 directional channels of GMDN \add{by using the atmospheric response of each detector to primary GCRs and the asymptotic look directions based on calculated GCR trajectories through the magnetosphere. The median rigidities of primary GCRs monitored by these detectors range over 11.3 - 106.9 GV, while the asymptotic look directions outside the magnetosphere cover almost the whole sky around Earth. The observed count rates are all corrected for atmospheric effects \citep[the atmospheric pressure effect for NM data and the atmospheric pressure and temperature effects for MD data;][]{Mendonca2016}} and normalized to the average count rate over the reference interval.\par
For each order of anisotropy ($n=0,1,2$), 
the intensity decrease \add{in space during each hour} is modeled as a power law in rigidity and
the \add{12} fit parameters are three
power-law indices $\gamma_n$ plus 
nine anisotropy amplitudes $\xi_c^{n,m}(t)$ and $\xi_s^{n,m}(t) $
($m=0,\dots,n$) at a \add{given} reference rigidity $P_r$.  
From the best-fit parameters, the fractional intensity change in space at rigidity $P$ and time $t$ relative to the intensity during the reference interval before the FD is given as
\begin{eqnarray}
\Delta I(P,t)&=&\sum_{n=0}^{2}\sum_{m=0}^{n} P_n^m(\theta) \{ \xi_c^{n,m} \cos m(\phi+\omega t)\nonumber \\
&&+\xi_s^{n,m}(t) \sin m(\phi+\omega t) \}(P/P_r)^{\gamma_{n}(t)}
\label{eq:DI}
\end{eqnarray}
where $P_n^m$ is the semi-normalized spherical function, 
$\omega=\pi/(12{\rm\ h})$, $\theta$ and $\phi$ are the geographical co-latitude and longitude of the asymptotic \add{look} direction \add{for} GCRs of rigidity $P$ outside the magnetosphere and $t$ is universal time in hours. We calculate $\Delta I(P,t)$ to be observed by AMS-02 by setting $P_r$ equal to the value $P_k=10.5$ GV mentioned above.  
For $P$ close to $P_r$, the GCR rigidity spectrum of the reference time interval $J(P)=J_0(P/P_r)^{-\mathit{\Gamma}(P)}$ is modulated at time $t$ during the FD to become 
$j(P,t)=J(P)[1+\Delta I(P,t)]=J_0[1+\Delta I(P_r,t)](P/P_r)^{-[\mathit{\Gamma}(P)+\Delta\mathit{\Gamma} (P,t)]}$. 
Using $\Delta I(P,t)$ from Eq.~(\ref{eq:DI}), the change in the spectral index at rigidity $P_r$ and time $t$ is given by $\Delta \mathit{\Gamma} (P_r,t)=-P_r(\partial/\partial P) \log[1+\Delta I(P,t)]|_{P=P_r}$.\par

\subsection{Neutron Monitor Leader Fraction Analysis}

\begin{figure}
\includegraphics[width=0.5\textwidth]{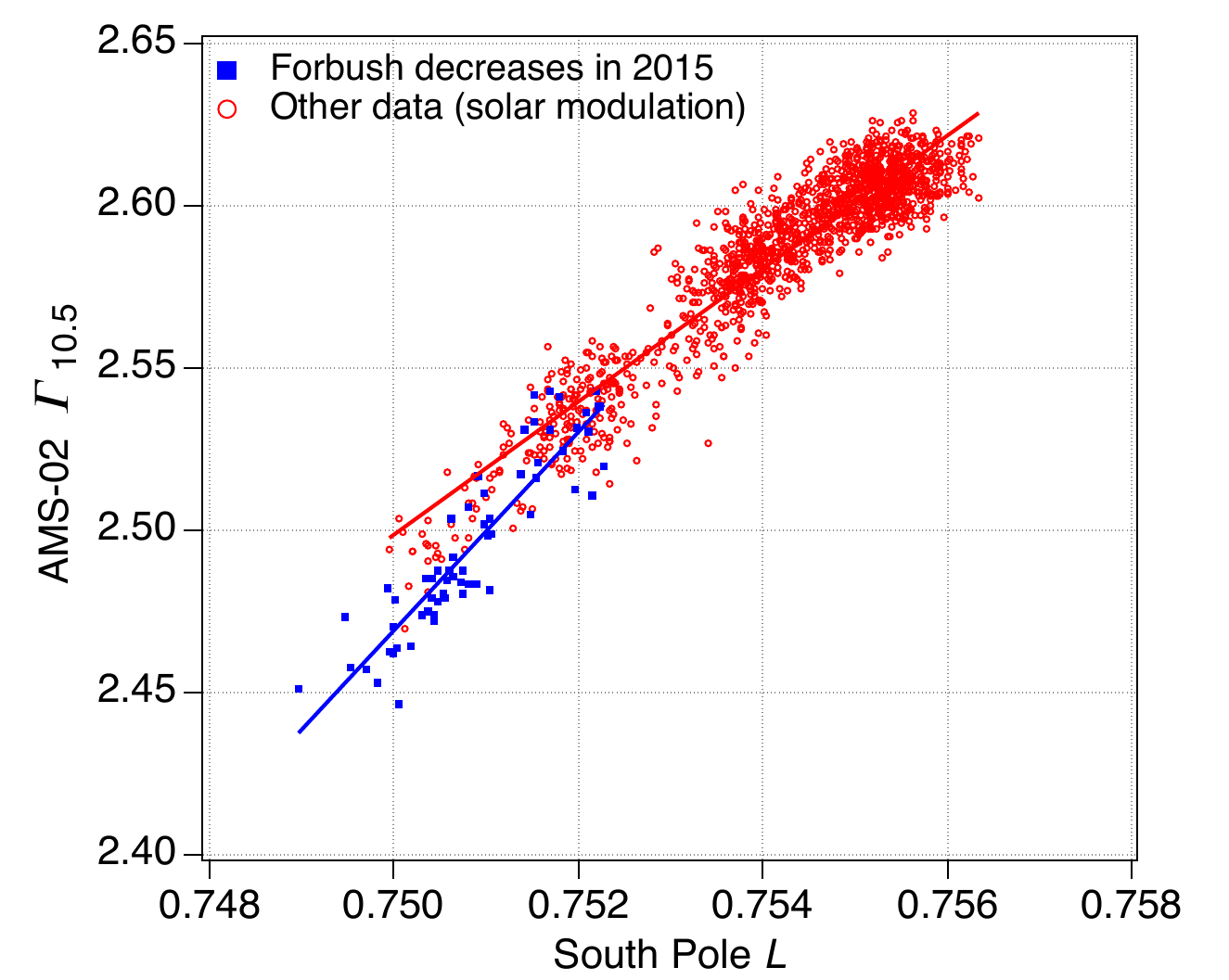}
\caption{Daily South Pole NM leader fraction $L$ versus AMS-02 $\mathit{\Gamma}_{10.5}$ (cosmic ray spectral index at 10.5~GV). Blue squares: the two FD periods in 2015.  
Red circles: all other data during 2015--2019, indicating variation with the solar activity cycle. 
Slopes of linear fits to the two data sets are $20.4\pm 0.2$ (red) and $30.7\pm 1.8$ (blue).}
\label{fig:LvsAMS}
\end{figure}

The NM leader fraction ($L$) \citep{Ruffolo16,Banglieng20} is defined as the fraction of detected neutrons that are ``leaders,'' i.e., not following other neutrons from the same primary cosmic ray. Using specialized electronics, we record distributions of time delays between two consecutive events detected by the same neutron counter \citep{Bieber04-mult}. We use a statistical method to calculate $L$ for each counter every hour by accumulating the time-delay ($\tau$) distribution which would follow an exponential function ($n\propto\exp[-\alpha \tau]$) for randomly unassociated events (leaders). The measured distribution is clearly higher than the exponential trend for short time delays, indicating that some events are temporally associated, i.e., there is a multiplicity of neutron counts from the same cosmic ray shower.  
The leader fraction $L$ is defined as 
\begin{equation}
    L=\frac{\rm 
    Integrated~counts~from~exponential~fit}{\rm Total~counts},
\label{eq:L}
\end{equation}
which represents the inverse multiplicity, and multiplicity relates to primary cosmic ray rigidity, so $L$ indicates the cosmic ray spectral index. 
Measurements of $L$ extend the capability of NMs to observe cosmic-ray spectral variation in addition to monitoring the count rate variation \citep{Ruffolo16,Mangeard16-latsur,Banglieng20}. 
In this work, we focus on
$L$ from the South Pole NM
because it
was calibrated vs.\ the spectral index inferred from AMS-02 daily proton data \citep{Muangha24}. 
Here we compare the South Pole $L$ \citep[corrected for pressure variation;][]{Muangha24} with the AMS-02 spectral index $\mathit{\Gamma}_{\rm 10.5}$ at 10.5~GV
for all times of overlapping data (Figure~\ref{fig:LvsAMS}). The correlation between $L$ and AMS-02 $\mathit{\Gamma}_{\rm 10.5}$ is well described by a linear fit with significantly different slopes during the two FDs of 2015 (blue) compared with all other data (red), which characterize solar modulation (i.e., variation with the solar activity cycle) during $2015-2019$.
Here we use the linear fit parameters for the two FD periods in 2015 (blue line 
with slope $30.7\pm1.8$) to infer \add{$\mathit{\Gamma}_{10.5}$,}
\add{and $\Delta\mathit{\Gamma}_{10.5}$ is defined as in Section 2.1.
The FD of 2017 is the remaining FD for which AMS-02 data are available; the data for this FD were not included among the blue points so that we can test whether the fit slope determined from the 2015 events is also applicable for this case.}

\begin{figure*}
    \includegraphics[width=0.5\textwidth]{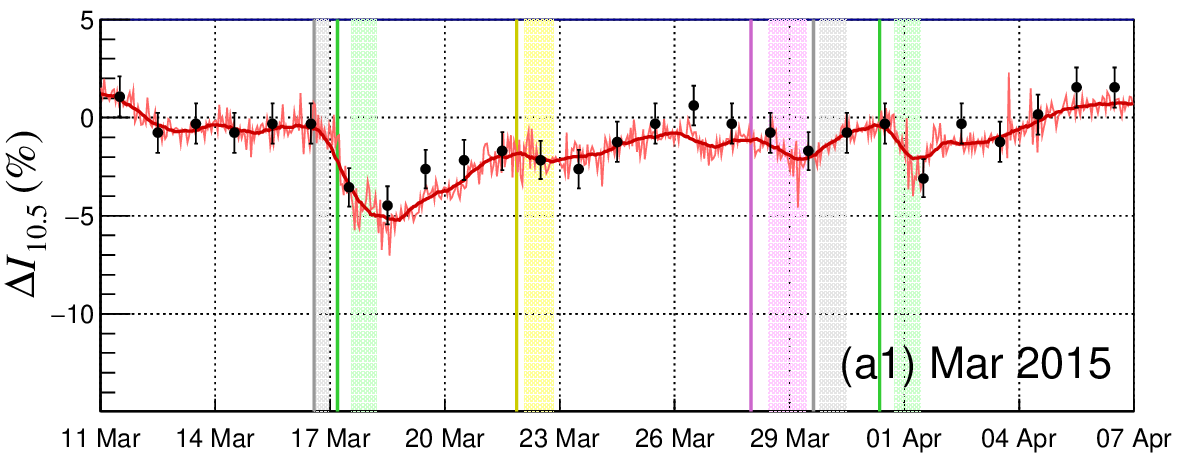}
    \includegraphics[width=0.5\textwidth]{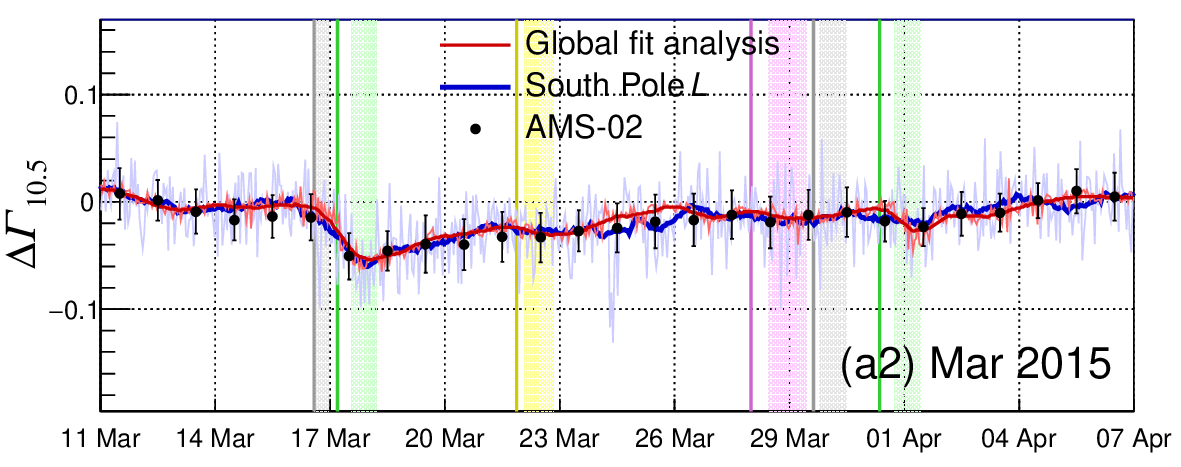}
    \includegraphics[width=0.5\textwidth]{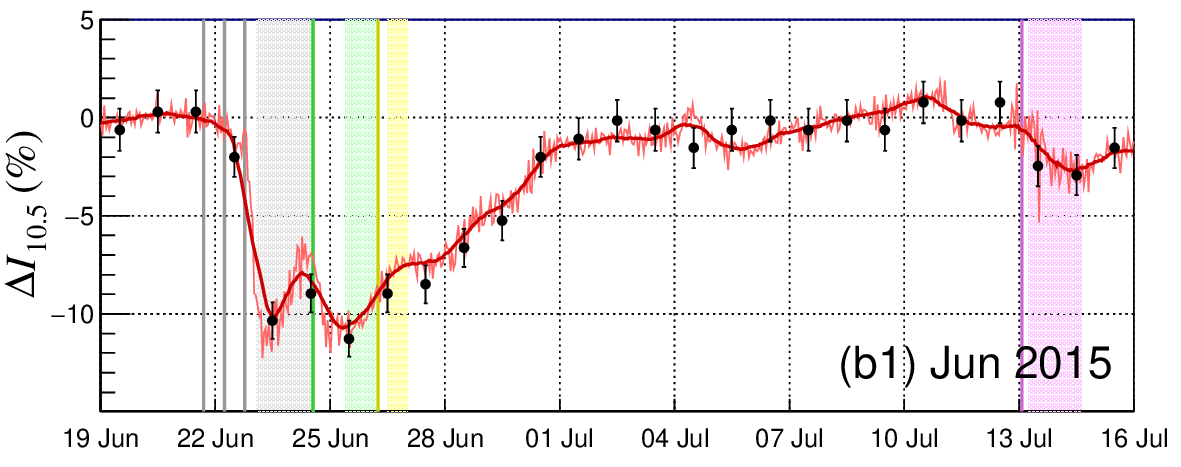}
    \includegraphics[width=0.5\textwidth]{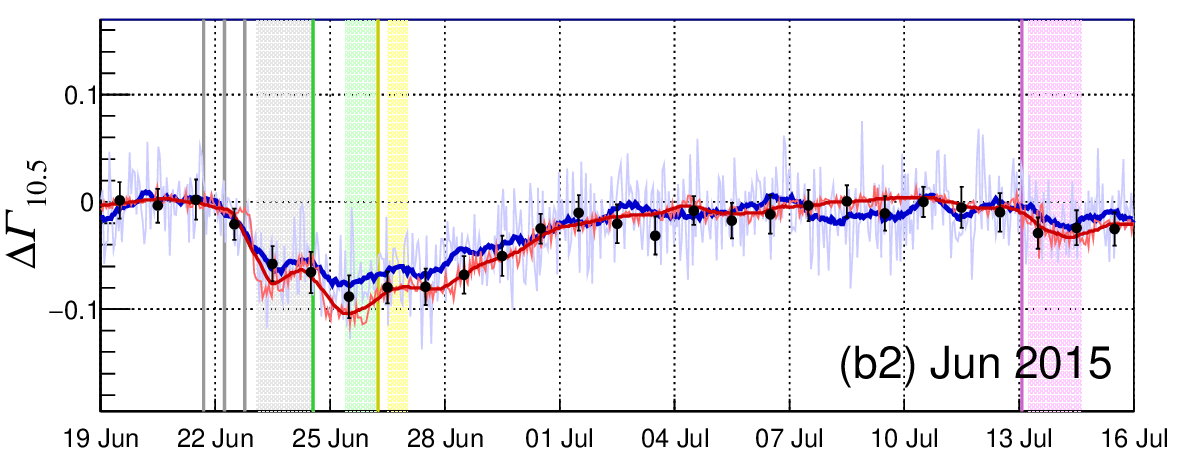}
    \includegraphics[width=0.5\textwidth]{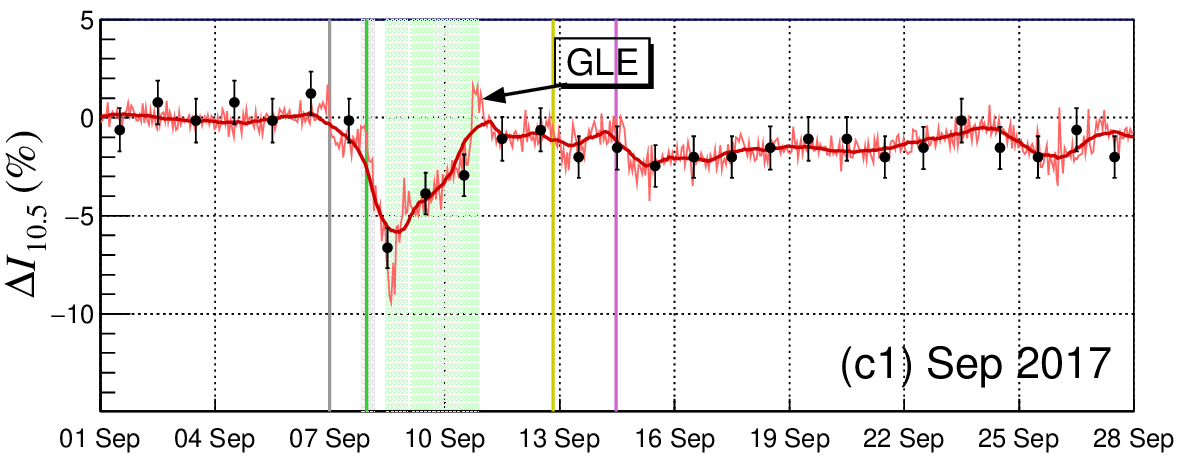}
    \includegraphics[width=0.5\textwidth]{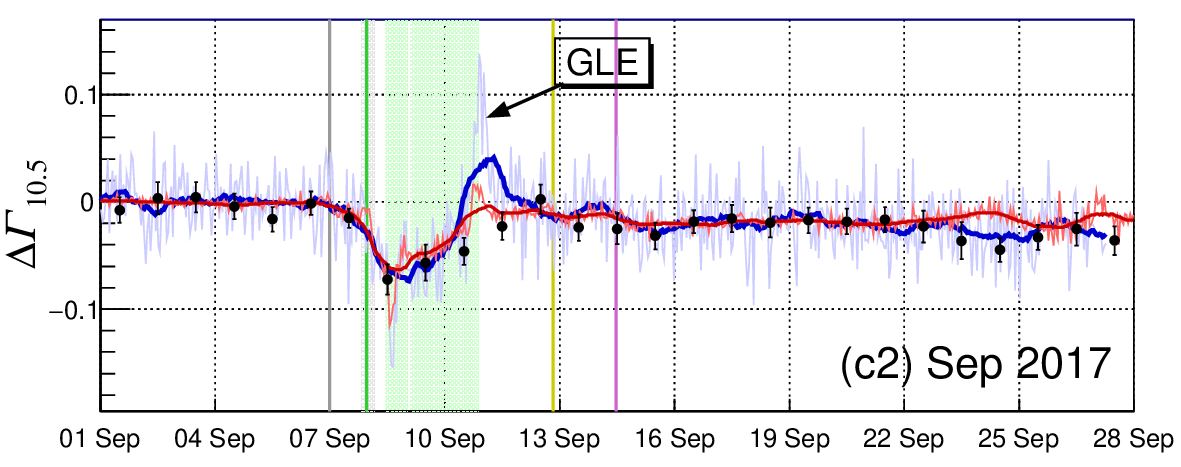}
    \includegraphics[width=0.5\textwidth]{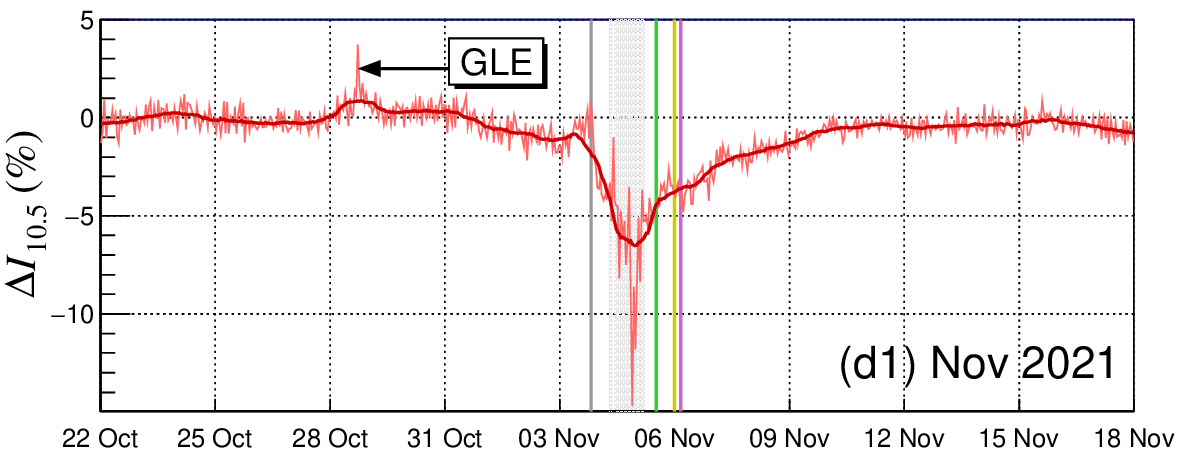}
    \includegraphics[width=0.5\textwidth]{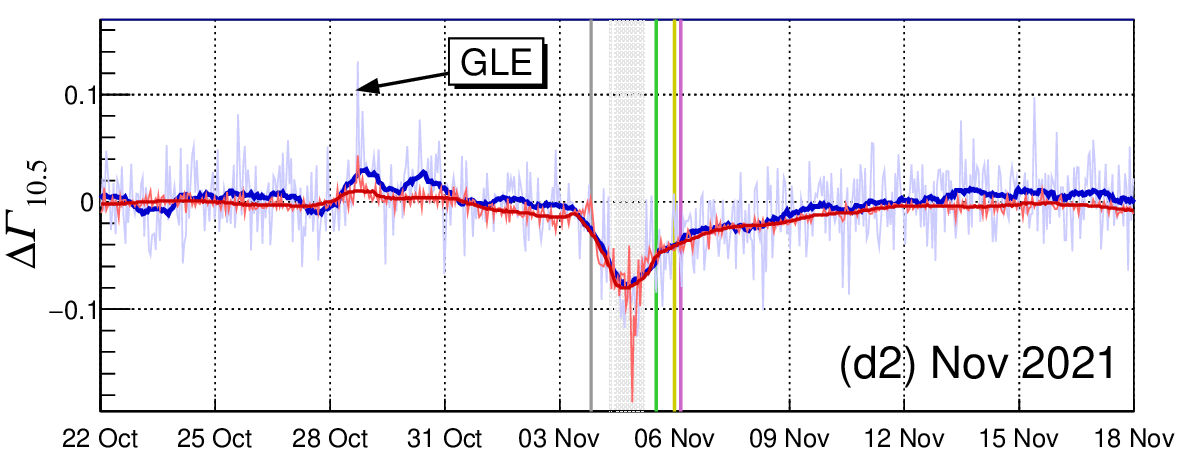}
    \includegraphics[width=0.5\textwidth]{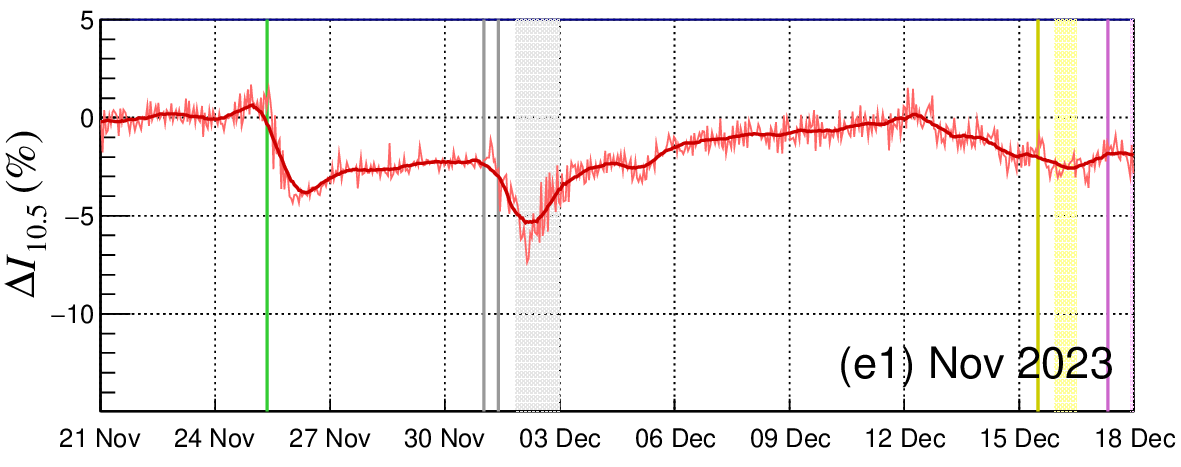}
    \includegraphics[width=0.5\textwidth]{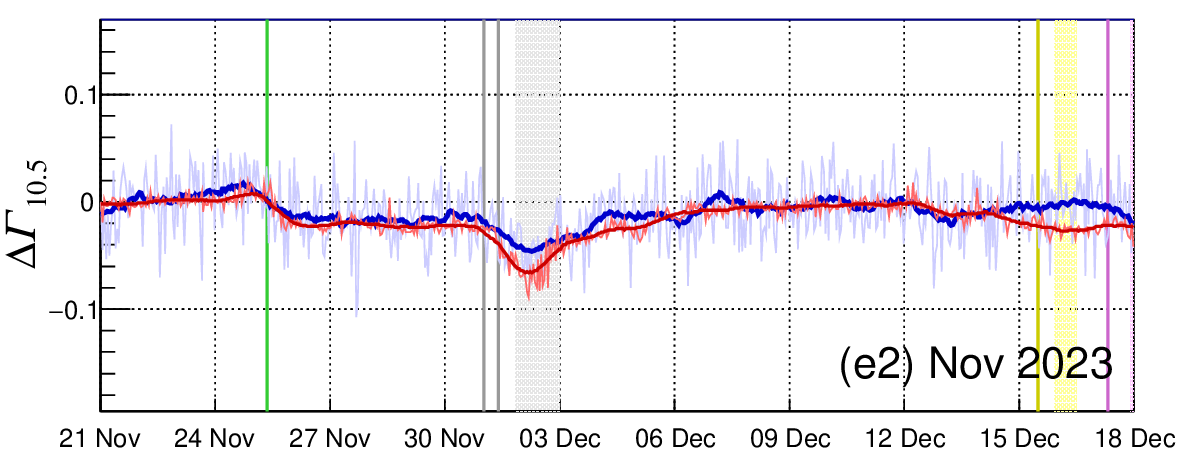}

\caption{
Temporal variations of cosmic ray flux $\Delta I(P,t)$ (left) and spectral index $\Delta \mathit{\Gamma}(P,t)$ (right) relative to a reference interval (first 3 days of each plot), based on two ground-based observing techniques, GFA (red) and South Pole $L$ (blue) [hourly data: thin, light curves; 24-hour running averages: thick, dark curves] and\add{, when available,} space-based daily AMS-02 proton data (black). 
Each panel displays values for rigidity 10.5 GV during 27-day solar rotation periods including five FDs starting in (a) March 2015, (b) June 2015, (c) September 2017, (d) November 2021, and (e) November 2023. 
FDs are associated with interplanetary shock arrival at Earth (vertical line) and/or passage of an ICME (shading of the same color).
There is very good consistency between all measures of GCR flux and spectral variations.
Two GLEs (solar energetic particle enhancements) in (c) September 2017 and (d) October 2021 were observed by ground-based detectors but excluded from published AMS-02 data.
}
\label{fig:FDs}
\end{figure*}

\section{Results} \label{sec:results}

Figure~\ref{fig:FDs} shows changes in the cosmic ray intensity $\Delta I(P_r,t)$ (left) and power-law index $\Delta \mathit{\Gamma}(P_r,t)$ (right) at $P_r=10.5$ GV over a synodic solar rotation period (27 days) including each of the five FDs.  
These quantities are differences relative to the reference interval (first three days in each plot). 
\add{In each panel, the hourly value from GFA and its 24-hour running average are shown by thin light red and thick dark red curves, respectively, while $\Delta \mathit{\Gamma}(P,t)$ derived from the hourly South Pole $L$ and its 24-hour running average are shown by thin light blue and thick dark blue curves, respectively.}  
Note that NM data are not shown in the left panels because a single NM does not measure $\Delta I$ at $P_r$; it responds to a wide range of primary GCR rigidity.
\add{For the three FDs in (a)-(c) for which the AMS-02 data are available, solid black circles show $\Delta \mathit{\Gamma}(P_r,t)$ as extracted from the AMS-02 observations.}

\add{Physically, FDs are associated with arrival of an interplanetary shock or passage of a driving ICME, so in Figure~\ref{fig:FDs} we have indicated these by a vertical line and with shading of the same color. \addd{The properties of the large ICMEs, our measured results, and geomagnetic disturbances are summarized in Table~\ref{tab:table1}, as described in the table caption.}
ICMEs and associated shocks were identified from a list by Richardson and Cane\footnote{https://doi.org/10.7910/DVN/C2MHTH}, and we also plotted shock arrival times from a list of sudden geomagnetic storm commencements (SSCs) as compiled by the International Service of Geomagnetic indices\footnote{http://isgi.unistra.fr}.}

Results from the three independent measurements of $\Delta\mathit{\Gamma}$ by AMS-02, GFA, and $L$ are in quite good \add{quantitative} agreement with each other at the same rigidity. \red{Many previous reports have analyzed the data observed by NMs and space-borne detectors, but usually only with qualitative or approximate comparisons of data for different rigidity ranges. Figure~\ref{fig:FDs} demonstrates, for the first time, the reliability and accuracy of ground-based observation of spectral variation. Note also that even hour-to-hour variation can be examined in the hourly GFA and $L$ results, while for AMS-02 only daily mean values have been made available. In this sense, the ground-based observations monitor GCR spectral variations with better temporal resolution.} \add{The measurement by AMS-02, on the other hand, provides us with the absolute flux as a function of rigidity, which is difficult to derive accurately from ground-based observations.}  

\red{The better time resolution in $L$ and GFA results is highlighted by the sharp increases due to ground level enhancements (GLEs) from solar energetic particles on 10 September 2017 and 28 October 2021 in $\Delta I(P,t)$ and $\Delta \mathit{\Gamma}(P,t)$ from GFA and South Pole $L$.  The GLE times were excluded from the published AMS-02 data.} \add{During these rare and brief GLE events (indicated in Figure~2) when} there is a mixture of solar and Galactic particles, which have very different spectra, $\Delta\mathit{\Gamma}$ is no longer a measure of the GCR spectral index, but rather a measure of effective spectral change that serves as an indicator of a GLE.
Note also that GLEs have very steep spectra so they more strongly affect the South Pole NM data at lower rigidity.

\begin{deluxetable*}{lrrcccll}
\tablewidth{0pt}
\tablecaption{\addd{Summary of five FDs analyzed in this work and their strongest associated ICMEs, from a list by Richardson and Cane$^1$. The shock arrival time ($t_s$) is rounded to the closest half hour. The ICME plasma start and end times and $\Delta T_{\rm max}$ values are expressed as offsets (in h) from $t_s$. Here $\Delta T_{\rm max}$ is the elapsed time from $t_s$ until the maximum decrease of the 24-hour running average of $\left|\Delta I_{\rm10.5}\right|$ or $\left|\Delta\mathit{\Gamma}_{\rm10.5}\right|$; maximum decrease values determined by GFA and $L$ are indicated. The uncertainties of ($\left|\Delta\mathit{\Gamma}_{\rm10.5}\right|\times$100) as estimated from the standard deviation of the 24-hour running average values during all of the first three reference days are 0.33 and 0.65 for GFA and $L$, respectively. The maximal planetary K-index ($K_p$) and auroral electrojet index (AE) for these events are also shown along with $\Delta T_{\rm max}$,  their offset time from $t_s$, in parentheses.} \label{tab:table1}}
\tablehead{
\colhead{ICME shock} &\multicolumn{2}{c}{ICME plasma (+h)} & \colhead{Maximum $\left|\Delta I_{\rm10.5}\right|$ (\%)} & \multicolumn{2}{c}{Maximum $\left|\Delta\mathit{\Gamma}_{\rm10.5}\right|\times$100} & \colhead{$K_p$} & \colhead{AE (nT)} \\
\colhead{arrival time ($t_s$)} & \colhead{Start} & \colhead{End} & \colhead{GFA (+$\Delta T_{\rm max}$)} & \colhead{GFA (+$\Delta T_{\rm max}$)} & \colhead{$L$ (+$\Delta T_{\rm max}$)} & \colhead{(+$\Delta T_{\rm max}$)} & \colhead{(+$\Delta T_{\rm max}$)}}
\startdata
2015/03/17 04:30 & +8.5 & +24.5 & 5.23 (+37) & 5.35 (+19) & 6.03 (+18) & 7.67 (+9) & 1570 (+10) \\
2015/06/22 18:30 & +7.5 & +43.5 & 10.7 (+62) & 10.4 (+62) & 7.83 (+66) & 8.33 (+1) & 1636 (+0)\\
2017/09/07 23:30 & +11.5 & +69.5 & 5.83 (+20) & 6.26 (+17) & 7.39 (+26) & 8.33 (+14) & 1442 (+15) \\
2021/11/03 19:30 & +11.5 & +33.5 & 6.51 (+28) & 7.99 (+19) & 7.75 (+22) & 7.67 (+15) & 1674 (+16) \\
2023/12/01 09:30 & +10.5 & +38.5  & 5.35 (+17) & 6.60 (+17) & 4.62 (+21) & 6.67 (+1) & \phantom{0}895 (+1) \\
\enddata
\tablecomments{\addd{The $K_p$ index is a 3-hour interval value from \url{https://kp.gfz.de/en/data}; the AE is an hourly value from \url{https://wdc.kugi.kyoto-u.ac.jp/aedir/.}}}
\end{deluxetable*}

\section{Discussion} \label{sec:discussion}

\add{When the interplanetary shock and the ICME embedded in the solar wind pass across Earth, the GCR density at Earth decreases. This FD is caused by physical processes including a diffusive barrier, closed magnetic field topology inside the CME, and adiabatic cooling in the expanding region  behind the shock \citep{Wibberenz98,KrittinathamRuffolo09,muna22}. The fractional decrease of GCR density due to adiabatic cooling is roughly independent of rigidity, but at higher rigidity, the GCRs from outside the FD refill the depleted region faster because of their longer mean free path (m.f.p.) of interplanetary scattering \citep{Engelbrecht22}. As a result, the magnitude of the observed density depression in a FD is smaller for GCRs at higher rigidity than at lower rigidity. Thus analyses of the rigidity spectrum of FDs enable us to deduce the rigidity dependence of the m.f.p., which reflects the power spectrum of the IMF fluctuation. By utilizing GFA of the data recorded by worldwide NMs and MDs, \citet{muna24} inferred the rigidity dependence of the m.f.p.\ as it changed dynamically during a FD event in response to Earth's magnetic connection along different field lines to different portions of the interplanetary shock passing away from Earth.}

Very intense geomagnetic storms, including substorm and auroral phenomena, are almost always associated with FDs \citep{Vennerstrom16}. 
However, the relationship is too complex for any one observed parameter to represent the severity of geomagnetic storms or the occurrence of substorms and
auroras.
For example, \citet{Belov24} found only a moderate correlation coefficient (up to 0.67) between the maximum change in cosmic ray flux during a FD (which in the present work is measured by $|\Delta I_{10.5}|$) and the intensity of an associated geomagnetic storm.
\red{Nevertheless, Table~\ref{tab:table1} shows a notable correlation between the changes in the cosmic ray flux spectral index, $|\Delta\mathit{\Gamma}_{10.5}|$, and the $K_p$ and AE values, especially when we derive $|\Delta\mathit{\Gamma}_{10.5}|$ from the South Pole leader fraction. If this correlation is supported by further work, our method could report the overall severity of an FD-associated geomagnetic storm in real time.}


\begin{figure}
\includegraphics[width=0.5\textwidth]{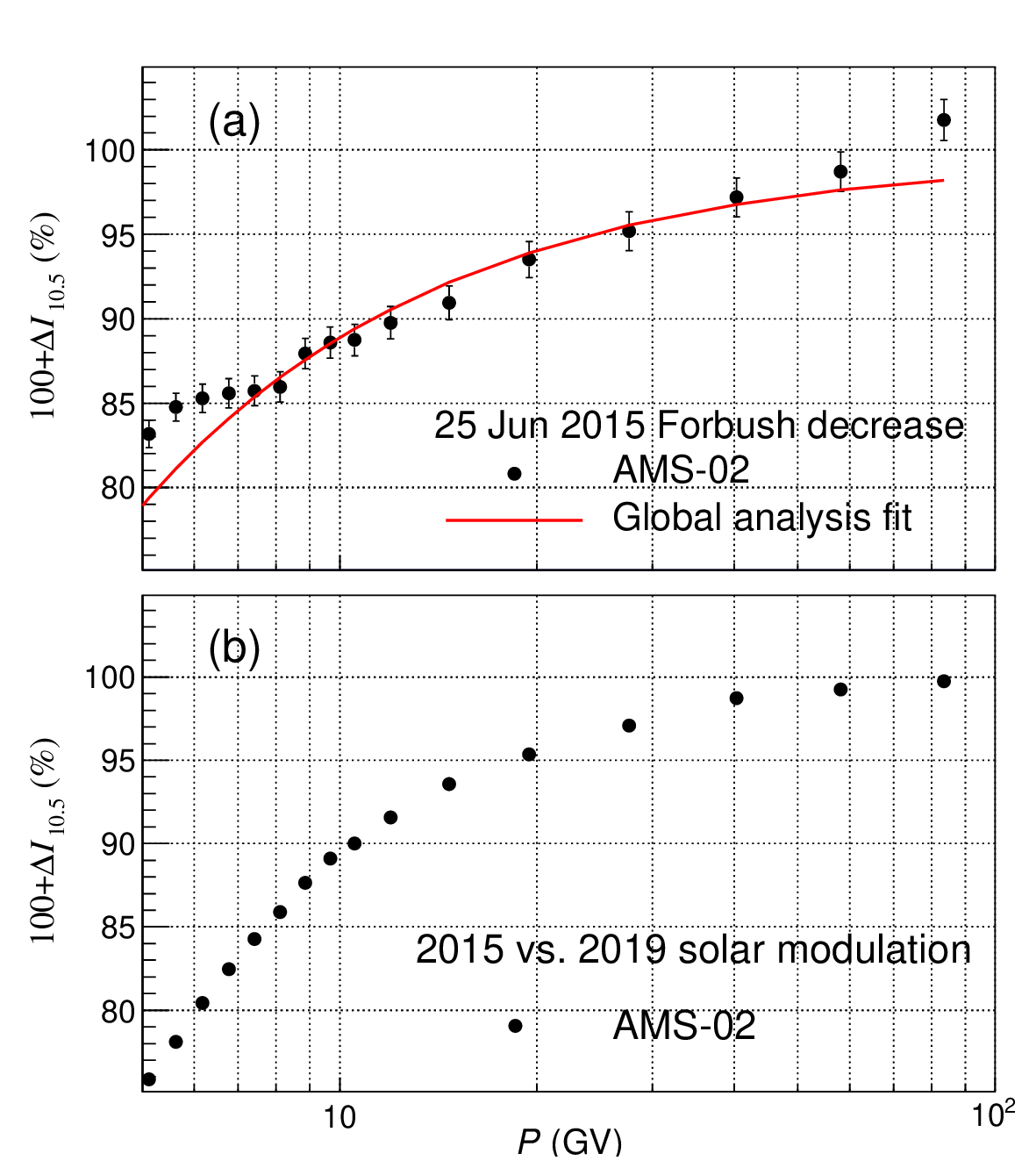}
\caption{
(a) Daily mean $\Delta I_{\rm 10.5}$ vs.\ rigidity observed by AMS-02 (black) and by GFA (red) on 25 June 2015, a time of maximum FD.
(b) $\Delta I_{\rm 10.5}$ vs.\ rigidity observed by AMS-02 (black) during 15 April to 11 May 2015 at solar maximum relative to 18 October to 13 November 2019 during solar minimum.
At $P\gtrsim7$ GV, both types of changes can be described by a power law deficit, while at lower rigidity the spectral changes of FDs and solar modulation are different.
}
\label{fig:spec}
\end{figure}

Analyzing the daily mean rigidity spectra observed by AMS-02 during a large number of FDs \add{in the previous solar cycle}, \citet{wang23} reported that the rigidity spectrum of the \add{fractional} GCR flux decrease, \add{which relates to the rigidity dependence of the m.f.p.,} is generally better described by an exponential function than by a power-law function. An example is shown in the left panel of Figure~\ref{fig:spec} which displays $\Delta I(P,t)$ on 25 June 2015, when the maximum GCR depression was recorded in the FD at 10.5 GV, as a function of rigidity. $\Delta I(P,t)$ from AMS-02 gradually decreases with decreasing rigidity down to $\sim$7 GV and then the spectrum becomes flatter. For physical discussions of this flattening and the temporal variation of the rigidity spectrum in FDs, see \citet{Alania12}, \citet{Munini18}, \citet{Belov21}, \citet{Blanco24}, and \citet{Mishev24}. 
However, \add{at much higher rigidity, where NMs and MDs precisely observe small but significant FD amplitudes, GFA results based on a fractional flux decrease with a power-law dependence on rigidity can effectively reproduce count rates observed by NMs and MDs \citep{muna22}. \citet{muna24} also reported a temporal change in the power-law index from GFA of NM and MD data during an extended decrease in January-February, 2012 known as a ``phantom FD."}
In Figure~3(a), the red curve is the fit result from the daily mean $\Delta I(P,t)$ by GFA (which includes muon detector channels with median rigidity over 100 GV), which cannot reproduce the flattening at low rigidity because it assumes a power-law spectrum over the entire rigidity range (see Eq.~\ref{eq:DI}), but matches well above $\sim$7 GV, where ground-based detectors are more sensitive.  
This is consistent with the good agreement between GFA and AMS-02 results at 10.5 GV in Figure~\ref{fig:FDs}(a-c). It is noted that the rigidity spectrum changes dynamically and its flattening also changes even in a single FD. \red{The measurement by AMS-02 enables us to investigate this flattening over a wide rigidity range,  including some of the higher rigidity range monitored by the ground-based observations.}

\red{We also note that the rigidity spectral change due to the solar cycle modulation shown in Figure~\ref{fig:spec}(b) is quite different. Solid circles in this panel show the Bartels rotation (BR) averages of GCR flux (in \%) observed by AMS-02 in BR2479 (15 April to 11 May 2015) during the solar maximum period relative to that in BR2540 (18 October to 13 November 2019) during the solar minimum period. This spectrum is roughly consistent with a power-law spectrum that exhibits no flattening in the lower rigidity region below 7 GV. 
Note also that the different trendlines shown in Figure~\ref{fig:LvsAMS} for AMS-02 $\mathit{\Gamma}_{10.5}$ vs.\ South Pole $L$ for solar modulation or during FDs can be attributed to the different spectral variations illustrated by the two panels of Figure~\ref{fig:spec}, noting that the South Pole NM has some sensitivity at rigidities down to $\sim$1~GV.}

For each time period in Figure~\ref{fig:FDs}, the observed decreases in $\Delta I$ and $\Delta\mathit{\Gamma}$ are associated with the arrival of an interplanetary shock (vertical line) followed by its driving ICME (time of passage indicated by shading of the same color), except for the decrease on 25 Nov 2023.  
\red{This event was associated with a shock but no ICME at Earth, and indeed the shock may have partially resulted from a solar wind stream interaction.
This is the only event for which we observe almost no change in spectral index, $\Delta\mathit{\Gamma}$, indicating a nearly rigidity-independent decrease in cosmic ray intensity.  
This is consistent with other observations that high-rigidity cosmic rays, with larger gyroradii, can remain unaffected by ICMEs, which are relatively small structures, while interplanetary shocks can have impacts that are nearly rigidity-independent \citep{Buatthaisong22}.}

\red{The right panels of Figure~\ref{fig:FDs} show very good agreement between independent results from the GFA and South Pole $L$, improving our confidence in both sets of ground-based results. 
Then measurements of $\Delta I(P,t)$ and $\Delta \mathit{\Gamma}(P,t)$ with fine temporal resolution can contribute to GLE alerts and reveal GCR responses to solar storms, for monitoring space weather.
To this end, we generate automated real-time plots of $L$ measurements at 1-minute, 1-hour, and 6-hour time scales to indicate the GCR spectral index (see \url{https://neutronm.bartol.udel.edu/realtime/southpole.html} and \url{https://neutronm.bartol.udel.edu/~pyle/json/SOPO_RTLF.html}). The GFA technique also can be performed at near real-time as a cross check.}

\blue{In summary, this work has introduced two ground-based techniques for measuring the GCR spectral index versus time during a Forbush decrease and shows that their results
are quite consistent with each other and with direct space-based measurements by AMS-02.
In particular, we demonstrate that our new methodologies
for ground-based observations are capable of precisely monitoring GCR
spectral variations with finer time resolution and in real time
during space weather events.}

\begin{center}
    ACKNOWLEDGMENTS
\end{center}
This work was supported in Thailand by the National Science and Technology Development Agency (NSTDA) and National Research Council (NRCT): High-Potential Research Team Grant Program (N42A650868), 
and from the NSRF via the Program Management Unit for Human Resources \& Institutional Development, Research and Innovation (B39G670013).
It is supported in part by JSPS KAKENHI Grant Number JP24K07068 and joint research programs of the National Institute of Polar Research (AJ1007), the Institute for Space-Earth Environmental Research (ISEE), Nagoya University, and the Institute for Cosmic Ray Research (ICRR), University of Tokyo in Japan, by NASA award 80NSSC24K0846,
and by the US NSF Award for Collaborative Research (2112437 and 2112439) and predecessor awards for the South Pole Neutron
Monitor operation. The observations with GMDN are supported by Nagoya University, INPE and UFSM, the Australian Antarctic Division and by project SP01/09 of the Research Administration of Kuwait University. We acknowledge the NMDB database (\url{http://www.nmdb.eu}), founded under the European Union's FP7 programme (contract no. 213007) for providing data. The GMDN data are available at \url{http://hdl.handle.net/10091/0002001448}.



\bibliography{FDindex}{}
\bibliographystyle{aasjournal}



\end{document}